\documentclass[aps,amsmath,amssymb,10pt]{revtex4}
\usepackage[latin9]{inputenc}
\usepackage{ifsym}
\usepackage{tipa}
\usepackage{amsmath}
\makeatletter

\def\HollowBox #1#2{{\dimen0=#1 \advance\dimen0 by -#2
       \dimen1=#1 \advance\dimen1 by #2
        \vrule height #1 depth #2 width #2
        \vrule height 0pt depth #2 width #1
        \llap{\vrule height #1 depth -\dimen0 width \dimen1} 
       \hskip -#2
       \vrule height #1 depth #2 width #2}}

\begin{document}

\title{Particle Content of Quadratic and $f (R_{\mu\nu\sigma \rho})$ Theories in $(A)dS$}

\author{Bayram Tekin}  
\email{btekin@metu.edu.tr}
\affiliation{Department of Physics,\\
             Middle East Technical University, 06800, Ankara, Turkey}
 
\date{\today}

\begin{abstract}

We perform a complete decoupling of the degrees of freedom of quadratic gravity and the generic $f (R_{\mu\nu\sigma \rho})$ theory about any one of their possible vacua, {\it i.e.} maximally symmetric  solution,  find the masses of the spin-2 and spin-0 modes in explicit forms. 

\end{abstract}

\maketitle

\section{Introduction \label{intro}}

The problem we shall address is simple to state: what is the perturbative particle spectrum of  the generic gravity theory defined by the action 
\begin{equation}
I=\int d^{n}x\,\sqrt{-g}\, f\left( g^{\mu \nu }, R_{\rho\sigma \mu \nu}\right),
\label{theory1}
\end{equation}
about any one of its possible maximally symmetric solution? Here, $f$ is assumed to be an analytic function of its arguments, the inverse metric and the Riemann tensor.  We shall consider spacetimes dimensions with $n \ge 3 $ here, but $n=2$ case as a pure $f(R)$ theory can also be included in the discussion with a redefinition of the cosmological constant to appear below. The action is assumed to be diffeomorphism invariant (at least up to a boundary term). Depending on the powers of the Riemann tensor, the theory has generically many maximally symmetric solutions which can be found once the function$f$ is given. For example, if the highest power is $N$, there are  generically $N$ vacua, modulo the assumption that the parameters of the theory satisfy 
certain constraints so that the effective cosmological constant of these vacua are real. In any case, for the discussion to follow, all we need is that the theory has at least one maximally symmetric solution, (anti)-de Sitter (A)dS  spacetime, with an effective cosmological constant $\Lambda$ which is generically nonzero. 

The actual identification of the particle content with explicit expressions for the masses in (A)dS is easier said than done as we shall work out in this work.   The particle content of quadratic gravity in (A)dS will play a major role here. It turns out that, rather surprisingly, even though quadratic gravity has been studied for a long time-it is almost as old as General Relativity!- its full perturbative particle content in (A)dS with {\it explicit} expressions for the masses have not been found.  Stelle,  in his groundbreaking works, \cite{Stelle1,Stelle2} gave the masses in four dimensional {\it flat} backgrounds. In the  next section more discussion on the literature will be given. In any case, we will need the particle spectrum of quadratic theory to answer the question posed above.  The connection of the particle content of quadratic gravity and the  theory  (\ref{theory1}) will be clear in a moment.  But first let us briefly argue why one would be interested in this theory. 

By now, it is no secret that General Relativity needs to be replaced by a, quantum-corrected, interim theory,  below the Planck scale, with a higher derivative theory, in general, having an action of the form $ I =\int d^n x\sqrt{-\left|g\right|} f(g_{\mu \nu}, R_{\mu \nu \sigma, \rho}, \nabla  R_{\mu \nu \sigma \rho}, ...)$ with many powers of the Riemann tensor, its covariant derivatives and contractions in a (most probably) diffeormophism invariant way. In addition, there might, of course, appear non-minimally coupled fields, especially scalar fields, directly taking part in gravitation, ruining the equivalence principle. For this case we have not much to say here, but assuming that gravity is solely described by a classical  pseudo-Riemannian spacetime which solves the field equations coming from a generic action, we can inquire the  particle content of the theory and the stability of a given solution. More specifically, one is usually interested in the linear stability of the maximally symmetric critical metrics $\bar{g}_{\mu \nu}$ (flat, de Sitter  or anti- de Sitter spaces) which are the potential vacua   in the absence of sources. As any physically viable theory should have a stable vacuum, this  puts constraints on the form of possible low energy quantum gravity theories.  

Moreover, recently, in \cite{Tekin1,Tekin2}, we have answered the following question in the affirmative : can one construct higher order metric-based gravity theories that has only a single massless spin-2 excitation (no other local  degrees of freedom) and a unique viable vacuum just like Einstein's theory ?  The theory obtained in these works is of the Born-Infeld type as the uniqueness of the viable vacuum is a very strong condition. More generally, one can search for higher derivative metric-based theories with only a massless spin-2 graviton in their spectrum about (A)dS vacua. Then, one cannot have derivatives of the Riemann tensor in the action, since generically these will yield extra degrees of freedom. Therefore, in this work we shall stick to the action of the form given in (\ref{theory1}) whose field equations are still fourth order and so generically have a massless spin-2, a massive spin-2 and a massive spin-0 excitation, just like quadratic gravity.  The connection between the spectra and vacua of quadratic gravity and (\ref{theory1}) are summarized in Section III.

\section{Full Particle Spectrum of  Quadratic Gravity in (A)dS }

As stated above, to identify the particle spectrum ({\it i.e.} calculate the masses) of generic $f\left( g^{\mu \nu }, R_{\rho\sigma \mu \nu}\right)$ theory, the best way is to first carry out the procedure for the quadratic gravity with the action 
 \begin{eqnarray}
I & = & \int d^{n}x\,\sqrt{-g}\left[\frac{1}{\kappa}\left(R-2\Lambda_{0}\right)+\alpha R^{2}+\beta R_{\mu\nu}^{^{2}}+\gamma\left(R_{\mu\nu\sigma\rho}^{2}-4R_{\mu\nu}^{2}+R^{2}\right)\right],\label{eq:Quadratic_action}\end{eqnarray}
whose source-free field equations read  \cite{Deser}
\begin{align}
\frac{1}{\kappa}\left(R_{\mu\nu}-\frac{1}{2}g_{\mu\nu}R+\Lambda_{0}g_{\mu\nu}\right)+2\alpha R\left(R_{\mu\nu}-\frac{1}{4}g_{\mu\nu}R\right)+\left(2\alpha+\beta\right)\left(g_{\mu\nu}\square-\nabla_{\mu}\nabla_{\nu}\right)R\nonumber \\
+2\gamma\left[RR_{\mu\nu}-2R_{\mu\sigma\nu\rho}R^{\sigma\rho}+R_{\mu\sigma\rho\tau}R_{\nu}^{\phantom{\nu}\sigma\rho\tau}-2R_{\mu\sigma}R_{\nu}^{\phantom{\nu}\sigma}-\frac{1}{4}g_{\mu\nu}\left(R_{\tau\lambda\sigma\rho}^{2}-4R_{\sigma\rho}^{2}+R^{2}\right)\right]\nonumber \\
+\beta\square\left(R_{\mu\nu}-\frac{1}{2}g_{\mu\nu}R\right)+2\beta\left(R_{\mu\sigma\nu\rho}-\frac{1}{4}g_{\mu\nu}R_{\sigma\rho}\right)R^{\sigma\rho} & =0.\label{fieldequations}\end{align}
We work with the mostly plus signature, therefore $\Lambda >0$ corresponds to the de-Sitter spacetime. 
Let $\bar{g}_{\mu \nu}$ denote a maximally symmetric solution whose curvatures are defined as
\begin{equation}
\bar{R}_{\mu \rho \nu \sigma}  = \frac{ 2 \Lambda}{ (n-1)(n-2)} \big ( \bar{g}_{\mu \nu}  \bar{g}_{\rho \sigma} -   \bar{g}_{\mu \sigma}  \bar{g}_{\rho \nu} \big ), \hskip 0.5 cm \bar{R}_{\mu \nu} = \frac{2 \Lambda}{n-2} \bar{g}_{\mu \nu}, \hskip 0.5 cm \bar{R} = \frac{ 2 n \Lambda}{ n-2}.
\end{equation}
Then for a vacuum, the field equations reduce to a quadratic equation that determines the effective cosmological constant:
\begin{equation}
\frac{\Lambda-\Lambda_{0}}{2\kappa}+k \Lambda^{2}=0,\qquad k \equiv\left(n\alpha+\beta\right)\frac{\left(n-4\right)}{\left(n-2\right)^{2}}+\gamma\frac{\left(n-3\right)\left(n-4\right)}{\left(n-1\right)\left(n-2\right)}.\label{quadratic}\end{equation}
As noted in the Introduction, for $\Lambda$ to be real, there is  a constraint on the parameters of the theory, but this will not be relevant to the ensuing discussion. We assume there is an (A)dS vacuum.   Then considering generic perturbations about this vacuum defined as  $h_{\mu\nu}\equiv g_{\mu\nu}-\bar{g}_{\mu\nu}$, one can show that the field equations at the linear order reduce to \cite{Deser}
\begin{equation}
c\,\mathcal{G}_{\mu\nu}^{L}+\left(2\alpha+\beta\right)\left(\bar{g}_{\mu\nu}\bar{\square}-\bar{\nabla}_{\mu}\bar{\nabla}_{\nu}+\frac{2\Lambda}{n-2}\bar{g}_{\mu\nu}\right)R^{L}+\beta\left(\bar{\square}\mathcal{G}_{\mu\nu}^{L}-\frac{2\Lambda}{n-1}\bar{g}_{\mu\nu}R^{L}\right)=0,\label{Linearized_eom}\end{equation}
where the constant $c$ in-front of the linearized Einstein tensor reads 
 \begin{equation}
c\equiv\frac{1}{\kappa}+\frac{4\Lambda n}{n-2}\alpha+\frac{4\Lambda}{n-1}\beta+\frac{4\Lambda\left(n-3\right)\left(n-4\right)}{\left(n-1\right)\left(n-2\right)}\gamma.\label{eq:c}\end{equation}
One cautionary remark is apt here: even though $1/c$ may appear like the effective Newton's constant in this theory, as we shall see in a moment in the action formulation, this is not really correct.  
A further term will be added to $c$ which will then yield the effective Newton's constant of the theory.  All the information about the particle content is in the linearized  fourth order equation  (\ref{Linearized_eom}), but it is clear that this is a complicated  coupled equation of physical degrees of freedom as well as gauge degrees of freedom.  We have to decouple the physical modes. The linearized version of the cosmological Einstein tensor  is defined as $\mathcal{G}_{\mu\nu}^{L} \equiv \Big(R_{\mu\nu}-\frac{1}{2} g_{\mu\nu}R +\Lambda g_{\mu\nu} \Big)^L$  which reads
 \[
\mathcal{G}_{\mu\nu}^{L}=R_{\mu\nu}^{L}-\frac{1}{2}\bar{g}_{\mu\nu}R^{L}-\frac{2\Lambda}{n-2}h_{\mu\nu}.\]
 where the  linearized Ricci tensor $R_{\mu\nu}^{L}$ and scalar
curvature $R^{L}=\left(g^{\mu\nu}R_{\mu\nu}\right)^{L}$ are given
as \[
R_{\mu\nu}^{L}=\frac{1}{2}\Big (\bar{\nabla}^{\sigma}\bar{\nabla}_{\mu}h_{\nu\sigma}+\bar{\nabla}^{\sigma}\bar{\nabla}_{\nu}h_{\mu\sigma}-\bar{\square}h_{\mu\nu}-\bar{\nabla}_{\mu}\bar{\nabla}_{\nu}h\Big),\qquad R^{L}=-\bar{\square}h+\bar{\nabla}^{\sigma}\bar{\nabla}^{\mu}h_{\sigma\mu}-\frac{2\Lambda}{n-2}h.\]
Let us  say a few words about what is already known in this theory: for flat spacetime, as we noted in four dimensions, 8 degrees of freedom were identified in \cite{Stelle1}.  For the (A)dS backgrounds, in \cite{Gullu}, the scattering amplitude  at tree-level  between two sources in this theory (augmented with a Fierz-Pauli mass term) was computed from which in principle one can read the masses from the poles, but as the Lichnerowicz Laplacian is used in that work, it is not easy to directly see all the masses, even though the computation is useful for unitarity and discontinuity analysis.  In \cite{Townsend}, for $n=3$, a specific combination of the quadratic 
terms ($8 \alpha + 3 \beta =0$) was considered which has a massive spin-2 excitation and the resulting theory is "New Massive Gravity" (NMG). In \cite{Deser_PRL}, Einstein-Hilbert piece is amputated from the  NMG and the resulting theory has a massless spin-2 excitation.   The most general version of quadratic gravity in $n=3$ was considered in \cite{Canonical}:  the mass spectrum was found  after a long computation. Specifically, $h_{\mu \nu}$  was decomposed into its irreducible parts and the massive spin-0 and massless spin-2 modes were decoupled to calculate the masses.  In \cite{Lu} and \cite{Critical}, four and $n$ dimensional versions of this theory for tuned parameters that eliminate the massive modes were studied, hence the "Critical Gravity" was obtained.  Of course what makes the computation rather tricky is the fact that the background spacetime is not flat and the curvature contributes to the masses of the particles.  Here  we remedy the gap on this and give a relatively concise derivation of the spectrum in $n$ dimensions for generic values of the parameters in (A)dS.  

Directly extending the quadratic gravity action up to $\mathcal{O}(h^2)$ is a very cumbersome task and it is rather a long exercise to put the final result in an explicitly gauge invariant form. Therefore, the best way to proceed is to use ''inverse" calculus of variations and get the action that yields the linearized field equations (\ref{Linearized_eom}): let that action be 
$I ( h^2) = \int d^n x \sqrt{-g}  {\cal L}_2$. Then the second order Lagrangian  is obtained  by first multiplying the linearized field equations by $-\frac{1}{2}h^{\mu \nu}$ and integrating the result over spacetime to arrive at (after dropping the boundary terms)
\begin{equation}
{\cal L}_2 = - \frac{1}{2} \Big ( c + \frac{ 2 \Lambda \beta}{(n-1)(n-2)} \Big ) h^{\mu \nu} \mathcal{G}_{\mu\nu}^L  + \beta \mathcal{G}_{\mu\nu}^L \mathcal{G}^{\mu\nu}_L + 
\Big ( \alpha + \frac{ \beta ( 4-n)}{4} \Big) R_L^2.
\label{orderh2}
\end{equation}
Up to a boundary term, this $\mathcal{O}(h^2)$ action is invariant under background diffeomorphisms  of the form $\delta_\xi h_{\mu \nu} = \bar\nabla_\mu  \xi_\nu  +\bar\nabla_\nu  \xi_\mu$ since both the linearized Einsten tensor and the linearized curvature scalar are gauge-invariant.  One can fix the gauge at this stage, but we shall proceed without a choice of gauge. The minus factor in front of the Einsteinian piece is important as it is chosen to give the correct kinetic energy for the massless spin-2 graviton. Or equivalently, if we couple the theory to matter, that is the correct sign, from which we can also identify the effective Newton's constant as 
\begin{equation}
\frac{1}{\kappa_\text{eff}}\equiv\frac{1}{\kappa}+\frac{4\Lambda ( n \alpha + \beta)}{n-2}+\frac{4\Lambda\left(n-3\right)\left(n-4\right)}{\left(n-1\right)\left(n-2\right)}\gamma,
\end{equation}
which has the earlier noted shift from the constant $c$. 
In what follows we shall make frequent use of integration by parts an the "Hermitian" property of the operator that defined as $\mathcal{G}_{\mu\nu}^L \equiv  (\mathcal{O} h)_{\mu\nu}$. 
So, not to clutter the notation, we work with Lagrangian but drop the boundary terms.  To be able to identify the physical modes, let us introduce two auxiliary fields $f_{\mu \nu} $ and $\varphi$ to recast the Lagrangian as
\begin{eqnarray}
{\cal L}_2= -\frac{1}{\kappa_\text{eff}} \Big( \frac{1}{2}h^{\mu\nu}+  f^{\mu \nu} \Big) {\cal G}^L_{\mu \nu}(h)  - \frac{1}{4 \beta \kappa_\text{eff}^2} \Big (f_{\mu \nu } f^{\mu \nu } - f^2 \Big) +\varphi R_L - \frac{b}{2} \varphi^2, 
\label{expanded_action3}
\end{eqnarray}
where  $ f \equiv \bar{g}^{\mu \nu} f_{\mu \nu}$ and  the constant $b$ is found as
\begin{equation}
b\equiv   \frac{2(n-1)}{ 4 \alpha(n-1) + \beta n}.
\end {equation}
So, integrating out the auxiliary fields  in (\ref{expanded_action3}) gives us back our original action (\ref{orderh2}). To get rid of the $\varphi R_L$ term let us define a new field $\tilde{f}^{\mu \nu}$ as 
\begin{equation}
f^{\mu \nu} = \tilde{f}^{\mu \nu}  - \frac{2\kappa_\text{eff}}{n-2} \varphi \,\bar{g}^{\mu \nu}, \hskip 0.5 cm f = \tilde{f} - \frac{2 n \kappa_\text{eff}}{n-2} \varphi,
\end{equation}
which then  reduces  (\ref{expanded_action3}) to  
\begin{eqnarray}
{\cal L}_2 &=& -\frac{1}{\kappa_\text{eff}} \Big( \frac{1}{2}h^{\mu\nu}+  \tilde{f}^{\mu \nu} \Big) {\cal G}^L_{\mu \nu}(h)  - \frac{1}{4 \beta \kappa_\text{eff}^2} \Big (\tilde{f}_{\mu \nu } \tilde f^{\mu \nu } - \tilde{f}^2 \Big)  \nonumber \\
&-&\frac{n-1}{(n-2) \beta \kappa_{\text{eff}}}\varphi \tilde{f} \ + \Big ( \frac{n(n-1)}{\beta (n-2)^2} - \frac{b}{2} \Big )\varphi^2.
\label{expanded_action4}
\end{eqnarray}
As $\varphi$ appears without derivatives, we can integrate it out  to arrive at 
\begin{eqnarray}
{\cal L}_2 &=& -\frac{1}{\kappa_\text{eff}} \Big( \frac{1}{2}h^{\mu\nu}+  \tilde{f}^{\mu \nu} \Big) {\cal G}^L_{\mu \nu}(h)  - \frac{1}{4 \beta \kappa_\text{eff}^2} \Big (\tilde{f}_{\mu \nu } \tilde f^{\mu \nu } - \tilde{f}^2 \Big)  - \frac{1}{4 \beta \kappa_\text{eff}^2} \xi \tilde{f}^2,
\label{expanded_action5}
\end{eqnarray}
where the constant  $\xi$ is given as 
\begin{equation}
\xi \equiv \frac{ 4 \alpha(n-1) + \beta n}{4(\alpha n + \beta)} . 
\end{equation}
A further field definition is needed to decouple $h_{\mu \nu}$ and $\tilde{f}_{\mu \nu}$. By inspection one observes that the following definition does the job 
\begin{equation}
h_{\mu \nu}  \equiv  \tilde{h}_{\mu \nu}  -\tilde f_{\mu \nu}.
\end{equation}
With this our second order Lagrangian reduces to the decoupled form
\begin{eqnarray}
{\cal L}_2 =-\frac{1}{2 \kappa_\text{eff}} h^{\mu\nu} {\cal G}^L_{\mu \nu}(h)   +\frac{1}{2 \kappa_\text{eff}}f^{\mu\nu} {\cal G}^L_{\mu \nu}(f )  - \frac{1}{4 \beta \kappa_\text{eff}^2} \Big ({f}_{\mu \nu } f^{\mu \nu } -{f}^2 \Big)  - \frac{1}{4 \beta \kappa_\text{eff}^2} \xi {f}^2,
\label{expanded_action6}
\end{eqnarray}
where we removed all the tildes for notational simplicity. As the first term is just the linearized Einstein theory with an effective Newton's constant, as long as  $\kappa_\text{eff} >0$, it describes a massless unitary spin-2 excitation, which is the Einsteinian mode, that is the massless graviton. Immediately, it is also clear that the second term has the wrong sign, so there will be a massive ghost.  It is also clear that when $\xi=0$, the $f^{\mu \nu}$ part is just the Fierz-Pauli massive gravity (with the wrong kinetic sign of course). For $\xi \ne 0$ as in our case, there is an additional massive mode which we have to decouple. To be able to read the masses, let us vary the action with respect to $f^{\mu \nu}$ to get
\begin{equation}
{\cal G}^L_{\mu \nu}(f )  - \frac{1}{2 \beta \kappa_\text{eff}} \Big ({f}_{\mu \nu } - \bar{g}_{\mu \nu} f \Big)  - \frac{1}{2 \beta \kappa_\text{eff}} \xi \bar{g}_{\mu \nu}f=0,
\label{fequation}
\end{equation}
whose trace yields
\begin{equation}
R_L(f) +  \frac{1}{(n-2)\beta \kappa_\text{eff}}\Big (1- n +n\xi \Big ) f =0. 
\end{equation}
${\cal G}^L_{\mu \nu}(f )$ satisfies the background Bianchi identity, hence double-divergence of (\ref{fequation}) yields 
\begin{equation}
(\xi-1) \bar \square f + \bar{\nabla}^\mu  \bar{\nabla}^\nu f_{\mu \nu} =0.
\end{equation}
Making use of this in the trace equation and using the definition of $R_L$, one arrives at a massive scalar wave equation satisfied by the trace of the $f$ field:
\begin{equation}
\Bigg( \xi  \bar \square + \frac{2 \Lambda}{ n-2} - \frac{ 1 - n + n\xi }{(n-2)\beta \kappa_\text{eff}} \Bigg) f = 0,
\end{equation}
from which we can read the mass of the scalar mode as 
\begin{equation}
m_s^2 =  - \frac{1}{\xi} \Bigg(  \frac{2 \Lambda}{ n-2} - \frac{ 1 - n + n\xi }{(n-2)\beta \kappa_\text{eff}} \Bigg),
\label{scalarmode}
\end{equation}
which of course decouples from the spectrum for the Fierz-Pauli tuning $\xi =0$. It is then easy to see that the trace-free part of (\ref{fequation}) yields the usual Fierz-Pauli massive graviton with the mass-square 
\begin{equation}
m_g^2 = - \frac{1}{\beta \kappa_\text{eff}}.
\label{ghost}
\end{equation} 
Let us summarize the particle content  of $n$-dimensional quadratic gravity in (A)dS : there is a unitary massless spin-2 mode, that is the usual graviton, there is a massive spin-zero mode whose mass-square is given as  (\ref{scalarmode}) which should satisfy the Breitenlohner-Freedman bound in AdS , namely $m_s^2 \ge \frac{ n-1}{2(n-2) }\Lambda$ to be non-tachyonic, and there is a massive spin-2 ghost with the mass-square given as (\ref{ghost}).  All together in $n$ dimensions the quadratic gravity has  $\frac{n(n-3)}{2} + \frac{(n+1)(n-2)}{2} +1 = n(n-2)$ degrees of freedom.  As concrete examples let us consider the three and the four dimensional cases.

{\bf{ $n=3$ :}} The masses of the spin-2 and spin-0 modes respectively read
\begin{equation}
m_g^2 = - \frac{1}{\kappa \beta} - 4\left ( 1 + 3\frac{\alpha}{\beta} \right) \Lambda , \hskip 1 cm m_s^2 =  \frac{1}{(8 \alpha + 3 \beta) \kappa} - 
\frac{ 4(3 \alpha + \beta)}{(8 \alpha + 3 \beta)}\Lambda,
\end{equation}
which are the same as the ones found with the canonical method in \cite{Canonical}. Note that altogether, these are the 3 degrees of freedom in 3 dimensions since there is no massless graviton in the generic theory.  On the other hand, the choice $8\alpha +3 \beta=0$ leads to the decoupling of the scalar mode, yielding the NMG theory \cite{Townsend} with a massive graviton. 

{\bf $n=4$:} The masses of the spin-2 and spin-0 modes respectively read
\begin{equation}
m_g^2 = - \frac{1}{\kappa \beta} - 2\left ( 1 + 4\frac{\alpha}{\beta} \right) \Lambda , \hskip 1 cm m_s^2 =  \frac{1}{2( 3\alpha + \beta) \kappa}.
\end{equation}
Together with the massless spin-2 graviton, these modes exhaust the 8 degrees of freedom whose flat space versions in four dimensions were given by Stelle \cite{Stelle1}. It is interesting to note that, four dimensions is rather unique in the sense that it is the only dimension for which the mass of the scalar field  is not shifted due to the cosmological constant. Also, it is clear that for $3\alpha + \beta=0$, the scalar mode decouples, which corresponds to the Weyl-square corrected Einstein's theory.  It is a little cumbersome-looking, but it pays to write the masses in generic $n$-dimensions in terms of the parameters of the Lagrangian: the massive spin-2 mode has the mass-square 
\begin{equation}
m_g^2 = -\frac{1}{\beta  \kappa }-4 \Lambda\frac{(n-1) (\beta +\alpha  n)+\gamma 
   (n-4) (n-3)}{\beta  (n-2) (n-1)},
\end{equation}
while the massive spin-0 has 
\begin{equation}
m_s^2 = \frac{n-2}{\kappa  (4 \alpha  (n-1)+\beta 
   n)}+ \frac{4 \Lambda (n-4)\Big ((n-1) (\beta +\alpha  n)+\gamma  (n-3) (n-2) \Big)}{(n-1)
   (n-2) (4 \alpha  (n-1)+\beta  n)},
\end{equation}
from which one can study various specific theories. For example, as $n \rightarrow \infty$, both masses remain intact.  For the pure Einstein-Gauss-Bonnet theory, they become infinite and decouple, leaving only the Einsteinian massless mode as expected, since the Einstein-Gauss-Bonnet theory is a second order theory.

As we have seen in the above construction, the massive spin-2 mode is a ghost, therefore as long as it is in the spectrum, the theory is problematic at the linear level. Namely, the vacuum is not stable against the copious production of these states which lower the energy.  For the tuned case of letting both masses go to infinity, one arrives at the "critical gravity" {\cite{Lu,Critical}} in AdS. But at exactly in this point of the parameter  space, there arise asymptotically non-AdS logarithmic modes \cite{Alishah,Gurses}. These solutions are of the wave type and they are valid both as exact and as perturbative solutions and as  perturbative modes, they are ghosts \cite{Porrati}. There is no consistent truncation of them that yields a nontrivial theory. There is an important digression that we would like to make here: in some theories, a perturbative solution cannot be obtained   from the linearization of an exact solution, a phenomenon called "linearization instability". If linearization instability exists, 
one might have a hope of obtaining a consistent theory as the log-modes can be truncated  as is the case in three dimensional chiral gravity \cite{wei,strominger,carlip}. This says that the dangerous perturbative  log-modes of chiral gravity do not come from the linearization of exact solutions. This is not the case in critical gravity as these modes are also exact solutions. 
Therefore, for $ n \ge 4$ one must have $\beta  =0$ to avoid the massive spin-2 ghost.  Let us now study the generic gravity.  

\section{Full Particle Spectrum of  $f(R_{\mu \nu \sigma \rho})$ gravity  in (A)dS }

The first thing to note is that taking the Lagrangian
density as  a function of the Riemann tensor with two up and two down indices is better as one can do away with the inverse metric: 
\begin{equation}
\mathcal{L}= f\left(R_{\rho\sigma}^{\mu\nu}\right).
\label{org}
\end{equation}
Now the usual route to the particle spectrum of this theory is again to find the ${\cal{O}}(h)^2$ expansion of this action about any one of its potential vacua $\bar{g}_{\mu \nu}$. But as we have shown in sufficient detail in \cite{tah1,tah2,tah3,tah4}, for this purpose and for finding the vacua of the theory, it is actually best to construct a quadratic action that has the same vacua and the spectrum is this theory.  Namely, we need to construct the following action 
\begin{equation}
f_{\text{quad-equal}}\left(R_{\rho\sigma}^{\mu\nu}\right)=\frac{1}{\kappa}\left(R-2\Lambda_{0}\right)+\alpha R^{2}+\beta R_{\nu}^{\mu}R_{\mu}^{\nu}+\gamma\left(R_{\rho\sigma}^{\mu\nu}R_{\mu\nu}^{\rho\sigma}-4R_{\nu}^{\mu}R_{\mu}^{\nu}+R^{2}\right),\label{equal_quad}
\end{equation}
whose vacua and degrees of freedom match the theory we want to explore (\ref{org}). Namely, we have to relate the parameters in this theory  to the values of the $f$ function and its derivatives.  As explained in detail in the above quoted works, this can be done by the following Taylor series expansion 
\begin{equation}
f_{\text{quad-equal}}\left(R_{\rho\sigma}^{\mu\nu}\right)\equiv\sum_{i=0}^{2}\frac{1}{i!}\left[\frac{\partial^{i}f}{\partial\left(R_{\rho\sigma}^{\mu\nu}\right)^{i}}\right]_{\bar{R}_{\rho\sigma}^{\mu\nu}}\left(R_{\rho\sigma}^{\mu\nu}-\bar{R}_{\rho\sigma}^{\mu\nu}\right)^{i}.\label{eq:f_quad-equal_Riemann}
\end{equation}
Therefore, given  $f\left(R_{\rho\sigma}^{\mu\nu}\right)$ defining the theory, and denoting the background Riemann tensor as  
\begin{equation}
\bar {R}_{\rho\sigma}^{\mu\nu} = \frac{2 \Lambda}{(n-1)(n-2)} \Big (\delta_{\rho}^{\mu}\delta_{\sigma}^{\nu}-\delta_{\sigma}^{\mu}\delta_{\rho}^{\nu} \Big),
\end{equation}
 one has to compute the following two derivatives and contractions 
\begin{align}
\left[\frac{\partial 
f}{\partial R_{\rho\sigma}^{\mu\nu}}\right]_{\bar{R}_{\rho\sigma}^{\mu\nu}}R_{\rho\sigma}^{\mu\nu} & \equiv\zeta R,\label{eq:First_order}\\
\frac{1}{2}\left[\frac{\partial^{2}f}{\partial R_{\rho\sigma}^{\mu\nu}\partial R_{\lambda\gamma}^{\alpha\beta}}\right]_{\bar{R}_{\rho\sigma}^{\mu\nu}}R_{\rho\sigma}^{\mu\nu}R_{\lambda\gamma}^{\alpha\beta} & \equiv\alpha R^{2}+\beta R_{\sigma}^{\lambda}R_{\lambda}^{\sigma}+\gamma\left(R_{\rho\sigma}^{\mu\nu}R_{\mu\nu}^{\rho\sigma}-4R_{\nu}^{\mu}R_{\mu}^{\nu}+R^{2}\right),\label{Second_order}
\end{align}
which determine the constants  $\zeta$, $\alpha$, $\beta$, $\gamma$.  The constant  $\zeta$ appears in the bare Newton's constant of the equivalent quadratic theory as  
\begin{equation}
\frac{1}{\kappa}  =\zeta-\left(\frac{4\Lambda}{n-2}\left(n\alpha+\beta\right)+\frac{4\Lambda\left(n-3\right)}{n-1}\gamma\right),\label{kappa}
\end{equation}
while the bare cosmological constant of the equivalent theory reads as (see {\cite{tah1,tah2,tah3,tah4} for further details)
\begin{equation}
\frac{\Lambda_{0}}{\kappa}  =-\frac{1}{2}f\left(\bar{R}_{\rho\sigma}^{\mu\nu}\right)+\frac{\Lambda n}{n-2}\zeta-\frac{2\Lambda^{2}n}{\left(n-2\right)^{2}}\left(n\alpha+\beta\right)-\frac{2\Lambda^{2}n\left(n-3\right)}{\left(n-1\right)\left(n-2\right)}\gamma.\label{Lambda_0}
\end{equation}
So, these parameters are sufficient to  determine the quadratic theory that has the same particle spectrum as the $f(R^{\mu \nu}_{ \sigma \rho})$ theory and since  we found the spectrum of the former, we can simply read the spectrum of the latter.  Given a generic  $f(R^{\mu \nu}_{ \sigma \rho})$, let us summarize  the recipe:  compute the first and second derivative with respected to  $ R^{\mu \nu}_{ \sigma \rho}$  and use (\ref{Second_order}) to obtain $\zeta$, $\alpha$, $\beta$ and $\gamma$.  Then use (\ref{kappa}) to determine $\kappa$.  Compute 
$f\left(\bar{R}_{\rho\sigma}^{\mu\nu}\right)$ and use (\ref{Lambda_0}) to determine $\Lambda_0$. Now, one can use (\ref{quadratic}) to to determine the possible effective cosmological constants.  Once this is done, the masses given in the previous section for quadratic gravity yield the masses of the massive spin-2 and massive spin-0 modes in the generic $f(R^{\mu \nu}_{ \sigma \rho})$ gravity. What is rather remarkable is that one actually has to do 3 basic computations: the value of the Lagrangian density, the first and second derivatives of the Lagrangian density with respect to the up-up down-down Riemann tensor evaluated at the (A)dS background.

\section{Conclusions }

Using auxiliary fields we have decoupled the free particle spectrum of general quadratic gravity in $n$ dimensional constant curvature backgrounds and  calculated the masses  of the massive spin-2 and and massive spin-0 modes whose special forms only have appeared before, even though quadratic gravity has been of interest for along time.   Then finding a quadratic action that has the same free particle spectrum and vacuum equation as the generic  $f(R^{\mu \nu}_{ \sigma \rho})$ gravity, we found the particle spectrum of the latter theory. 
 Therefore once an explicit form of the action is given, no matter how complicated the action is, as long as it depends on the powers of the Riemann tensor (and its contractions, the Ricci tensor and the scalar curvature)  our formulas give the masses of the gravitons about the (A)dS backgrounds. Perturbative stability of vacuum of the generic theory is similar to the quadratic case that we have discussed in the text: the massive spin-2 mode is a ghost hence it should not appear in the spectrum.  We have given an example of a non-trivial theory in the Born-Infeld form  in \cite{Tekin1,Tekin2} which does not also have the spin-0 mode, see also another recent example \cite{Pablo}.

\begin{acknowledgments}

This work is partially supported by the Scientific and Technological Research 
Council of Turkey (T{\"U}B\.{I}TAK) through the grant 113F155. I would like to thank S. Deser, M. Gurses and T. C. Sisman for useful discussions on generic gravity theories and W. Li, W. Song and A. Strominger for an explanation of  a particular point of their chiral gravity work. 

\end{acknowledgments}


\begin{thebibliography}{99}

\bibitem{Stelle1} 
  K.~S.~Stelle,
 ``Renormalization of Higher Derivative Quantum Gravity,''
  Phys.\ Rev.\ D {\bf 16}, 953 (1977).


\bibitem{Stelle2} 
  K.~S.~Stelle,
 ``Classical Gravity with Higher Derivatives,''
  Gen.\ Rel.\ Grav.\  {\bf 9}, 353 (1978).


\bibitem{Tekin1} 
  I.~Gullu, T.~C.~Sisman and B.~Tekin,
 ``Born-Infeld Gravity with a Massless Graviton in Four Dimensions,''
  Phys.\ Rev.\ D {\bf 91}, no. 4, 044007 (2015).

\bibitem{Tekin2} 
  I.~Gullu, T.~C.~Sisman and B.~Tekin,
 ``Born-Infeld Gravity with a Unique Vacuum and a Massless Graviton,''
  Phys.\ Rev.\ D {\bf 92}, no. 10, 104014 (2015).


\bibitem{Deser}   S. Deser and B. Tekin,
``Gravitational energy in quadratic curvature gravities,''
  Phys. Rev. Lett. {\bf 89}, 101101 (2002);
 ``Energy in generic higher curvature gravity theories,''
  Phys. Rev. D {\bf 67}, 084009 (2003).

 \bibitem{Gullu} 
  I.~Gullu and B.~Tekin,
  ``Massive Higher Derivative Gravity in $D$-dimensional Anti-de Sitter Spacetimes,''
  Phys.\ Rev.\ D {\bf 80}, 064033 (2009).

\bibitem{Townsend} 
  E.~A.~Bergshoeff, O.~Hohm and P.~K.~Townsend,
 ``Massive Gravity in Three Dimensions,''
  Phys.\ Rev.\ Lett.\  {\bf 102}, 201301 (2009).

\bibitem{Deser_PRL} 
  S.~Deser,
 ``Ghost-free, finite, fourth order $D=3$ (alas) gravity,''
  Phys.\ Rev.\ Lett.\  {\bf 103}, 101302 (2009).

\bibitem{Canonical} 
  I.~Gullu, T.~C.~Sisman and B.~Tekin,
 ``Canonical Structure of Higher Derivative Gravity in 3D,''
  Phys.\ Rev.\ D {\bf 81}, 104017 (2010).

\bibitem{Lu} 
  H.~Lu and C.~N.~Pope,
 ``Critical Gravity in Four Dimensions,''
  Phys.\ Rev.\ Lett.\  {\bf 106}, 181302 (2011).

\bibitem{Critical} 
  S.~Deser, H.~Liu, H.~Lu, C.~N.~Pope, T.~C.~Sisman and B.~Tekin,
 ``Critical Points of D-Dimensional Extended Gravities,''
  Phys.\ Rev.\ D {\bf 83}, 061502 (2011).

\bibitem{Alishah} 
  M.~Alishahiha and R.~Fareghbal,
  ``$D$-Dimensional Log Gravity,''
  Phys.\ Rev.\ D {\bf 83}, 084052 (2011).

\bibitem{Gurses} 
  I.~Gullu, M.~Gurses, T.~C.~Sisman and B.~Tekin,
 ``AdS Waves as Exact Solutions to Quadratic Gravity,''
  Phys.\ Rev.\ D {\bf 83}, 084015 (2011).

\bibitem{Porrati} 
  M.~Porrati and M.~M.~Roberts,
 ``Ghosts of Critical Gravity,''
  Phys.\ Rev.\ D {\bf 84}, 024013 (2011).


\bibitem{wei} 
  W.~Li, W.~Song and A.~Strominger,
 ``Chiral Gravity in Three Dimensions,''
  JHEP {\bf 0804}, 082 (2008).

\bibitem{strominger} 
  A.~Maloney, W.~Song and A.~Strominger,
 ``Chiral Gravity, Log Gravity and Extremal CFT,''
  Phys.\ Rev.\ D {\bf 81}, 064007 (2010).

\bibitem{carlip} 
  S.~Carlip,
 ``Chiral Topologically Massive Gravity and Extremal B-F Scalars,''
  JHEP {\bf 0909}, 083 (2009).


\bibitem{tah1}  I.~Gullu, T.~C.~Sisman and B.~Tekin, ``Unitarity
analysis of general Born-Infeld gravity theories,'' Phys.\ Rev.\ D
\textbf{82}, 124023 (2010).

\bibitem{tah2} I.~Gullu, T.~C.~Sisman and B.~Tekin,
``All Bulk and Boundary Unitary Cubic Curvature Theories in
Three Dimensions,'' Phys.\ Rev.\ D \textbf{83}, 024033 (2011).

\bibitem{tah3} T.~C.~Sisman, I.~Gullu and B.~Tekin,
``All unitary cubic curvature gravities in D dimensions,''
Class.\ Quant.\ Grav.\ \textbf{28}, 195004 (2011).


\bibitem{tah4} 
  C.~Senturk, T.~C.~Sisman and B.~Tekin,
 ``Energy and Angular Momentum in Generic F(Riemann) Theories,''
  Phys.\ Rev.\ D {\bf 86}, 124030 (2012).

\bibitem{Pablo} 
  P.~Bueno, P.~A.~Cano, A.~O.~Lasso and P.~F.~Ramirez,
 ``f(Lovelock) theories of gravity,''
  arXiv:1602.07310 [hep-th].  

\end{thebibliography}
\end{document}